%% file: main.tex
\documentclass{llncs}

\newcommand{\ExtendedVersionOnly}[1]{#1}\newcommand{\ProceedingsVersionOnly}[1]{}

\usepackage{algorithm}
\usepackage{alltt}
\usepackage{amsfonts}
\usepackage{amsmath}
\usepackage{amssymb}
\usepackage[USenglish]{babel}
\usepackage{booktabs}
\usepackage{color}
\usepackage{eepic}
\usepackage{epic}
\usepackage{epsfig}
\usepackage{etex}
\usepackage{fancyvrb,relsize}
\usepackage{graphicx}
\usepackage[colorlinks=true,linkcolor=black,citecolor=black]{hyperref}
\usepackage{multicol}
\usepackage{multirow}
\usepackage{paralist}
\usepackage{stmaryrd}
\usepackage{subfigure}
\usepackage{todonotes}
\usepackage{url} 
\usepackage[all]{xy}
\ExtendedVersionOnly{\usepackage{times}}

\hypersetup{bookmarksopen=true}
\input{inc_draft}
\input{inc_symbols}
\input{inc_terminology}

\begin{document}
\allowdisplaybreaks

\ProceedingsVersionOnly{
	\title{Bindings-Restricted Triple Pattern Fragments}
	\nocite{ExtendedVersion}
}

\ExtendedVersionOnly{
	\title{brTPF: Bindings-Restricted Triple Pattern Fragments}
	\subtitle{\vspace{-1mm}(Extended Version)%
	\vspace{-6mm}\footnote{This document is an extended preprint of a paper published in the proceedings of the ODBASE~2016 conference~\cite{ProceedingsVersion}. In contrast to the proceedings version, this document contains Appendixes~\ref{appendix:Virtuoso} and~\ref{appendix:QET} which present additional experimental results.}}
}

\author{%
	Olaf Hartig\inst{1,2}
	\and
	Carlos Buil-Aranda\inst{3}
}
\institute{
		Hasso Plattner Institute,
		University of Potsdam,
		Germany\\
	\and
		Dept.\ of Computer and Information Science (IDA),
		Link\"oping University,
		Sweden\\
		\email{olaf.hartig@liu.se}
		\\[2mm] 
	\and
		Informatics Department,
		Universidad T\'ecnica Federico Santa Mar\'ia,
		Chile\\
      \email{cbuil@inf.utfsm.cl}
}

\sloppy

\maketitle

\ExtendedVersionOnly{ \vspace{-5mm} } 
\begin{abstract}
\input{abstract}
\end{abstract}

\ExtendedVersionOnly{ \vspace{-7mm} } 
\input{introduction}

\input{RelatedWork}

\input{formalization}

%
%
%
%

\input{ImplementationDescription}
\input{NetworkLoadExperiments}
\input{ServerLoadExperiments}
\input{CacheExperiments}

\input{Conclusions}

\medskip\noindent\textbf{Acknowledgements:}
Olaf Hartig's work has been funded partially by the German Government, Federal Ministry of Education and Research
	\ProceedingsVersionOnly{under the project number 03WKCJ4D.}
	\ExtendedVersionOnly{(project no.~03WKCJ4D).}
Carlos Buil-Aranda's work has been supported by the Millennium Nucleus Center for Semantic Web Research
	\ProceedingsVersionOnly{under Grant NC120004 and by UTFSM DGIP Project no. 116.24.1}
	\ExtendedVersionOnly{(Grant NC120004) and by UTFSM DGIP Project no.~116.24.1}

\ExtendedVersionOnly{\enlargethispage{\baselineskip}} 
\ExtendedVersionOnly{ \vspace{-1mm} } 

\bibliographystyle{splncs03}
\bibliography{references}

\ExtendedVersionOnly{

\appendix

%


\newpage
\noindent \textbf{\large Appendix} \vspace{-5mm}
\input{AppendixVirtuoso}
\input{AppendixQueryExecTimes}

}  

\end{document}

%% file: inc_draft.tex
\definecolor{darkgray}{rgb}{.3,.3,.3}

\newcommand{\FinalVersion}[1]{#1}\newcommand{\DraftVersion}[1]{}

\FinalVersion{\newcommand{\removable}[1]{#1}}
\DraftVersion{\newcommand{\removable}[1]{{\color{darkgray}#1}}}

%% file: inc_symbols.tex
\newcommand{\definedAs}    
	{=}

\newcommand{\fctsymDom}{\mathrm{dom}} 
\newcommand{\fctDom}[1]{\fctsymDom(#1)}

\newcommand{\tuple}[1]{\langle #1 \rangle} 

\newcommand{\symURI}{u} 

\newcommand{\symRDFgraph}{G} 

\newcommand{\symTP}{tp} 

\newcommand{\fctsymControl}{c}
\newcommand{\fctControl}[1]{\fctsymControl(#1)}
\newcommand{\symControls}{C}

\newcommand{\fctsymSelector}{s}
\newcommand{\fctSelector}[1]{\fctsymSelector(#1)}
\newcommand{\fctsymBJTPSelector}[2]{\mathrm{s}_{(#1,#2)}}
\newcommand{\fctBJTPSelector}[3]{\fctsymBJTPSelector{#1}{#2}(#3)}

\newcommand{\symLDf}{f}
\newcommand{\symLDfElmts}{\Gamma}

\newcommand{\symLDfMeta}{M}

\newcommand{\symCollection}{F}


%% file: inc_terminology.tex
\newcommand{\brTPFnonabbrev}{bindings restricted triple pattern fragment}
\newcommand{\brTPF}{brTPF}
\newcommand{\brTPFs}{{\brTPF}s}

\newcommand{\definedTerm}[1]{\textbf{#1}}

\newcommand{\maxMpR}{\textsf{\footnotesize maxMpR}}

\newcommand{\numreq}{\textsf{\footnotesize \#req}}
\newcommand{\dataRecv}{\textsf{\footnotesize dataRecv}}
\newcommand{\throughput}{\textsf{\footnotesize throughput}}

\newcommand{\numhits}{\textsf{\footnotesize \#hits}}

%% file: abstract.tex
The Triple Pattern Fragment (TPF) interface is a recent proposal for reducing server load in Web-based approaches to execute SPARQL queries over public RDF datasets. The price for less overloaded servers is a higher client-side load and a substantial increase in network load (in terms of both the number of HTTP requests and data transfer).
In this paper, we propose a slightly extended interface that allows clients to attach intermediate results to triple pattern requests. The response to such a request is expected to contain triples from the underlying dataset that do not only match the given triple pattern (as in the case of TPF), but that are guaranteed to contribute in a join with the given intermediate result.
Our hypothesis is that a distributed query execution using this extended interface can reduce the network load (in comparison to a pure TPF-based query execution) without reducing the overall throughput
	\removable{of the client-ser\-ver system}
significantly.
Our main contribution in this paper is twofold: we empirically verify the hypothesis and provide an extensive experimental comparison of our proposal and TPF.

%% file: introduction.tex
\section{Introduction}
\label{sec:intro}

\ExtendedVersionOnly{ \vspace{-1mm} } 

Recent years have witnessed a large and constant growth in the amount of data that is structured based on the data model of the Resource Description Framework~(RDF)~\cite{Cyganiak14:RDFConcepts} and made available on the Web through HTTP interfaces~\cite{Bizer13:DeploymentOfStructuredDataOnTheWeb,Mika12:MetadataStatistics,Schmachtenberg14:AdoptionOfLinkedDataPractices}. A prevalent\removable{~(and standardized)} type of such interfaces that provides query-based access to RDF data are SPARQL endpoints~\cite{SPARQLprot}; that is, Web services that accept queries written in the SPARQL query language~\cite{Harris13:SPARQL1_1Language}. While a SPARQL endpoint enables users to query its RDF dataset by using the full potential of SPARQL, providing such a comparably complex functionality presents a serious challenge~(the evaluation problem of a core fragment of SPARQL has been shown to be PSPACE complete~\cite{SCSPARQL}). As a consequence, many public endpoints suffer from frequent downtime; for instance, by monitoring over 400 such endpoint for 27 months, Buil-Aranda et al.\ show that only 32.3\% of the endpoints offer an availability of more than 99\%, and 50.4\% have an availability of less than 95\%~\cite{Buil-Aranda:2013:SWI:2717178.2717198}. Furthermore, if many client applications start to access
	such an
endpoint concurrently, then the performance of the endpoint~(in terms of average query execution times and per-client query throughput) drops significantly~\cite{Verborgh14:LDF}.

To address these problems, Verborgh et al.\ recently proposed the Triple Pattern Fragment~(TPF) interface~\cite{Verborgh14:LDF,Verborgh16:LinkedDataFragmentsArticle}. This proposal restricts the type of queries supported by the server to single triple patterns. Then,
	to support arbitrary SPARQL queries over a dataset exposed by such a TPF server, a major part of the query processing effort has to be shifted to the clients.
As a result, the server load is reduced and query execution times are more stable. However,
	\removable{everything comes at a cost and, thus,}
the price for
	the \removable{aforementioned} benefits of the TPF approach
is not only a higher client-side load but also a significant increase in network load%
	. \removable{More precisely, to execute a given SPARQL query, a TPF-based client usually sends many more HTTP requests than a client that sends the whole query with a single request to a SPARQL endpoint. Additionally, the overall amount of data returned in response to these TPF requests is much greater than what a SPARQL endpoint returns~(namely, just the query~result).}

To mitigate this drawback of the TPF approach~(without losing the
	benefits)
we propose a slightly extended interface that supports so-called \emph{Bin\-dings-Re\-strict\-ed Triple Pattern Fragments}~(\emph{brTPF}). That is, in addition to pure TPF requests, the brTPF interface allows clients to attach intermediate results to TPF requests. The response to such a brTPF request is expected to contain RDF triples from the underlying dataset that do not only match the given triple pattern~(as in the case of TPF), but that are guaranteed to contribute in a join with the given intermediate result. Hence, given the brTPF interface,
	it becomes possible to distribute the execution of joins
between client and server by using the well-known bind join~strategy~\cite{Haas97:QueriesAcrossDiverseSources}.
Our hypothesis is
	given as follows: \par \smallskip \noindent \textbf{Hypothesis.} ~\textit{In comparison to pure TPF-based query executions, query executions that are based on the brTPF interface can reduce the network load without reducing the overall throughput \removable{of the client-ser\-ver system} significantly.} \par \smallskip
Our main contribution in this paper is twofold: First, we empirically verify the
	aforementioned
hypothesis. Second, we also provide an extensive experimental comparison of the brTPF approach and the TPF approach.
The queries that we focus on in this study are expressed using SPARQL basic graph patterns~(BGPs). We chose this focus because we believe that achieving a comprehensive understanding of how TPF and brTPF behave for this fundamental fragment of SPARQL is essential before attempting to focus on more expressive fragments. 
All digital artifacts related to our study~(e.g., software, test data, etc.) are available
	\ProceedingsVersionOnly{on the Web page for this paper.\footnote{\url{http://olafhartig.de/brTPF-ODBASE2016}}}%
	\ExtendedVersionOnly{on-line at: {\footnotesize \url{http://olafhartig.de/brTPF-ODBASE2016}}.}

The remainder of the paper is organized as follows:
Section~\ref{sec:related_work} provides an overview of the related work on existing HTTP interfaces to access RDF data on the Web.
Section~\ref{sec:formalization} presents a formalization of the concepts used in this paper, Section~\ref{sec:ImplementationDescription} describes the implementation of the propotypes for the brTPF server and client, as well as a brief description of the TPF  server and client implementations. Sections~\ref{sec:NetworkLoadExperiments},~\ref{sec:ServerLoadExperiments} and~\ref{sec:CacheExperiments} present the experiments that we have done to evaluate TPF and brTPF in terms of the network load, throughput, and cache use, respectively. Finally, we conclude in Section~\ref{sec:conclusions}. 
%

%% file: RelatedWork.tex
\section{Related Work}
\label{sec:related_work}

\begin{figure}[t]
\centering
%
%
\includegraphics[width=\textwidth]{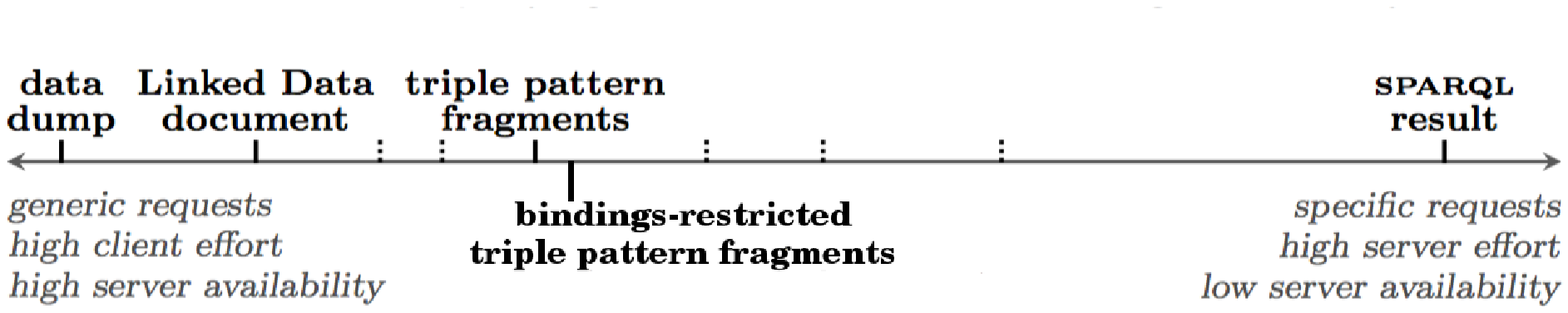}%
\vspace{-2mm} 
\caption{\footnotesize HTTP interfaces to RDF data~(adapted from~\cite{Verborgh14:LDF}).}
\label{fig:tpf_spectrum}
\vspace{-3mm} 
\end{figure}

\ProceedingsVersionOnly{
Verborgh et al.\ classified HTTP interfaces for accessing RDF data on the Web along an axis as illustrated in Figure~\ref{fig:tpf_spectrum}~\cite{Verborgh14:LDF}. In the following we describe the currently most common of these~interfaces.
}

\ExtendedVersionOnly{
One of the most common ways for accessing RDF data on the Web is through HTTP interfaces.
	Verborgh et al.\ classified such interfaces along an axis as illustrated in Figure~\ref{fig:tpf_spectrum}~\cite{Verborgh14:LDF}. In the following we describe the currently most common of these~interfaces.
}


\subsection{SPARQL Endpoints}
SPARQL endpoints are Web services that implement the SPARQL Protocol~\cite{SPARQLprot}. A SPARQL endpoint is usually an HTTP interface that accepts SPARQL queries (like Jena Fuseki\footnote{\url{https://jena.apache.org/documentation/serving_data/}}) and the query processing is done by the endpoint's triple store~(e.g., Virtuoso\footnote{\url{http://virtuoso.openlinksw.com/}}, Jena TDB\footnote{\url{https://jena.apache.org/documentation/tdb/}}). As it was shown in~\cite{Buil-Aranda:2013:SWI:2717178.2717198}, public SPARQL endpoints
may suffer from low availability and poor performance.

\subsection{Triple Pattern Fragments}

Verborgh et al.~\cite{Verborgh14:LDF} proposed the \emph{Triple Pattern Fragments}~(TPF) interface to access RDF data which relies on client processing power to execute SPARQL queries while the server only provides the data for the triple patterns in the queries. By relying on the clients to execute the actual operations on the data, it is possible to
	provide higher availability with the same server infrastructure,
since servers only perform operations that require minimal effort.
\ProceedingsVersionOnly{ 

}%
The originally proposed client-side algorithm for the TPF interface is based on a decomposition of a given query into triple patterns%
	; the resulting subqueries are then executed recursively in dynamically generated pipelines that order the subqueries based on result size metadata as provided by the TPF server%
~\cite{Verborgh14:LDF,Verborgh16:LinkedDataFragmentsArticle}. In comparison to SPARQL endpoints, the downside of this approach is not only a higher client-side load but also an increase in network load~(in terms of both the number of HTTP requests and data~transfer).

To
	reduce the impact of this downside
some new approaches to
	execute SPARQL queries over the TPF interface
appeared.
	For instance, Van Herwegen et al.\
extend the original TPF client algorithm by injecting every partial solution into the next subquery with which it shares variables~\cite{DBLP:conf/esws/HerwegenVMW15}.
%
Acosta et al.~\cite{DBLP:conf/semweb/AcostaV15}
	introduce a new TPF client algorithm using a query optimizer that
focuses on reducing both the amount of intermediate results and the requests posed to the TPF server, and it also implements an adaptive routing query engine that is able to dynamically adapt the query plan according to execution conditions. These works focus only on optimizing the client algorithm while the server remains unchanged, and even though the approaches improve TPF-based query execution considerably, we argue that there is room for improvement at the server side. 

There have been proposals to extend the TPF server instead of focusing on the clients. Van Herwegen et al.\ extended the original TPF controller to deal with substring filtering~\cite{DBLP:conf/semweb/HerwegenVVMW15}. The authors extended the server capabilities using an Elastic Search service, allowing the server to execute FILTER queries besides the usual triple patterns. Another approach, by Vander Sande et al., is to extend the TPF metadata by using approximate membership functions~(Bloom filters and Golomb-coded sets)~\cite{DBLP:conf/semweb/SandeVHMW15}. As a result of using these statistical techniques, the amount of HTTP requests can be reduced by~25\%, barely increasing the server load. On the other hand, due to an increased response size, query execution times could not be improved by these techniques.

In addition to a single-server scenario, it has also been shown that distributed query processing is possible over a federation of multiple TPF servers~\cite{Verborgh16:LinkedDataFragmentsArticle}.
	In this context,
Montoya et al.\ introduce a TPF-based federation system that is aware of replicated data~\cite{DBLP:conf/semweb/MontoyaSMV15}. Given a set of TPF servers with data replicated across these servers, the system creates an efficient execution plan for a given query by identifying which fragments will contribute to the final query result. Again, this work focuses mainly on the client, and tries to minimize the amount of data transferred from the servers to the client.

\subsection{Linked Data Querying}

Hartig et al.~\cite{hartigetal} proposed a Linked Data traversal approach, in which the follow your nose concept is applied for following links between linked RDF datasets on the Web. The implementation consists first in a source selection process for next traversing the links obtained from the initial query to the selected data sources. The link traversal operator implements an asynchronous pipeline of iterators executing first the most selective iterator. The query engine is also able to adapt the execution to source availability by detecting whenever an HTTP server stops responding. This is one of RDF data access approaches that imposes less load on the server, however it is among the slowest.

%% file: formalization.tex
\section{Formal Definition of \brTPF} \label{sec:formalization}

In this section, we provide a formal definition of brTPF. We assume that the reader is familiar with the fundamental concepts of RDF~\cite{Cyganiak14:RDFConcepts} and SPARQL~\cite{Harris13:SPARQL1_1Language,SCSPARQL}.

Then, for our formalization we adopt the general formal framework for defining interfaces to RDF datasets as provided by Verborgh et al.~\cite{Verborgh16:LinkedDataFragmentsArticle}. As the basis for formalizing any type of such an interface, the authors introduce the notion of a \emph{selector}
	\emph{function}
that captures conditions for selecting sets of triples from datasets~(cf.~\cite[Definition~1]{Verborgh16:LinkedDataFragmentsArticle}).
	Formally, any selector function is a mapping between RDF graphs. \par
We define such a selector
	function
for \brTPFs\ as follows.

\begin{definition} \label{def:brTPselector}
	Given a triple pattern $\symTP$ and a finite sequence of solution mappings $\Omega$, the \definedTerm{bindings-restricted triple pattern selector} for $\symTP$ and $\Omega$, denoted by $\fctsymBJTPSelector{\symTP}{\Omega}$, is the selector function that, for every RDF graph $\symRDFgraph$, is defined~by
	\begin{equation*}
		\fctBJTPSelector{\symTP}{\Omega}{\symRDFgraph}
		=
		\begin{cases}
			\lbrace
				t \in \symRDFgraph
				\,|\,
				\text{$t$ is a matching triple for $\symTP$}
			\rbrace
				&\text{if $\Omega$ is empty,}
			\\[1mm]
			\lbrace
				t \in \symRDFgraph
				\,|\,
				\text{$t$ is a matching triple for $\symTP$, and there}
			& \text{else.}
			\\[-1mm]
				\hspace{12.5mm}
				\text{exists a solution mapping $\mu$ such that}
			&
			\\[-1mm]
				\hspace{12.5mm}
				\text{the application of $\mu$ to $\symTP$ results in $t$}
			&
			\\[-1mm]
				\hspace{12.5mm}
				\text{and $\mu$ is compatible with some $\mu'$ in $\Omega$}
			\rbrace
			&
		\end{cases}
	\end{equation*}
\end{definition}

Given the general notion of a selector function, in \cite[Definition~3]{Verborgh16:LinkedDataFragmentsArticle}, Verborgh et~al.\ define the abstract concept of a \emph{Linked Data Fragment}~(LDF) of an RDF graph $\symRDFgraph$ as a tuple $\symLDf = \tuple{\symURI, \fctsymSelector, \symLDfElmts, \symLDfMeta, \symControls}$ where
$\symURI$ is a~URI~(from which $\symLDf$ can be retrieved);
$\fctsymSelector$ is a~selector function;
$\symLDfElmts$ is a~set of (blank-node-free) RDF triples that is the result of applying the selector function~$\fctsymSelector$ to $\symRDFgraph$, i.e., $\symLDfElmts \definedAs \fctSelector{\symRDFgraph}$;
$\symLDfMeta$ is a~finite set of (additional) RDF triples, including triples that represent metadata for~$\symLDf$;
$\symControls$ is a~finite set of hypermedia controls~(cf.\ \cite[Definition~2]{Verborgh16:LinkedDataFragmentsArticle}).
Additionally, by \cite[Definition~5]{Verborgh16:LinkedDataFragmentsArticle}, for a hypermedia control $\fctsymControl$~(cf.\ \cite[Definition~2]{Verborgh16:LinkedDataFragmentsArticle}), a \emph{$\fctsymControl$-specific LDF collection} over an RDF graph $\symRDFgraph$ is a~set~$\symCollection$ of LDFs such that, for each LDF $\symLDf \in \symCollection$ with $\symLDf = \tuple{\symURI, \fctsymSelector, \symLDfElmts, \symLDfMeta, \symControls}$, the following three properties hold:
i) $\symLDf$ is an LDF of $\symRDFgraph$,
ii) $\fctsymSelector \in \fctDom{\fctsymControl}$, and
iii) $\fctControl{\fctsymSelector} = \symURI$.

By using these preliminaries, we can now define \brTPFs\ as follows.


\begin{definition} \label{def:brTPF}
	Given
		a positive integer \maxMpR\ and
	a control $\fctsymControl$~(as per~\cite[Definition~2]{Verborgh16:LinkedDataFragmentsArticle}), a~$\fctsymControl$-specific LDF collection~$\symCollection$~(cf.\ \cite[Definition~5]{Verborgh16:LinkedDataFragmentsArticle}) is called a \definedTerm{\brTPFnonabbrev\ collection} for \maxMpR\ if, for any possible triple pattern $\symTP$ and any finite sequence $\Omega$ of
		at most \maxMpR\ distinct
	solution mappings, there exists an LDF $\tuple{\symURI, \fctsymSelector, \symLDfElmts, \symLDfMeta, \symControls} \in \symCollection$~(as per~\cite[Definition~3]{Verborgh16:LinkedDataFragmentsArticle}), referred to as a \definedTerm{\brTPFnonabbrev}~(\definedTerm{\brTPF}), that has the following three properties:
	\begin{enumerate}
		\item \label{item:bjTPF:1}
			Selector $\fctsymSelector$ is the bindings-restricted triple pattern selector for $\symTP$ and $\Omega$;
		\item \label{item:bjTPF:2}
			There exists a~(metadata) RDF triple $\tuple{\symURI,\mathtt{void\!:\!triples},cnt} \in \symLDfMeta$ with $cnt$ representing an estimate of the cardinality of $\symLDfElmts$; that is, $cnt$ is an integer that has the following two properties:
			\begin{enumerate}
				\item If $\symLDfElmts = \emptyset$, then $cnt \!=\! 0$.
				\item If $\symLDfElmts \neq \emptyset$, then $cnt \!>\! 0$ and $\mathrm{abs}\bigl( \left|\symLDfElmts\right| \!-\! cnt \bigr) \!\leq\! \epsilon$ for some $\symCollection$-specific threshold~$\epsilon$.
			\end{enumerate}
		\item \label{item:bjTPF:3}
			$\fctsymControl \in \symControls$.
	\end{enumerate}
\end{definition}

Observe that our definition assumes a positive integer \maxMpR\ that presents a well-defined restriction on the number of distinct solution mappings that can be attached to any brTPF request supported by any specific brTPF interface.
	Furthermore, note
that we require $\Omega$ to be a \emph{sequence} of solution mappings~(rather than a set). This requirement reduces the complexity of
	a correct~(i.e., deterministic) implementation of paging for the brTPF interface.

%% file: ImplementationDescription.tex
\section{Evaluation Prototypes}
\label{sec:ImplementationDescription}

This section describes the implementations that we used for our experiments.

\subsection{Server Implementation}

As a basis for the server, we used an established Java servlet implementation of the TPF interface\footnote{\url{https://github.com/LinkedDataFragments/Server.Java}} and extended it with the functionality to also support the brTPF interface. As a result, both interface implementations coexist within the Java servlet, which choses which of them it invokes depending on the HTTP GET request it receives: If the request contains a bindings-restricted triple pattern selector, then the brTPF implementation is used to generate the response; if the HTTP request just contains a TPF selector, the TPF implementation is used. Having both implementations in a single software component has the advantage that commonly used basic functionality~(such as serializing RDF triples for a response) is shared and, thus, experimental results are not affected by potential differences in how efficient the implementation of such basic functionality is.

\paragraph{TPF Server Implementation:}
The TPF server implementation is a Java servlet that accepts HTTP GET and POST requests. The servlet identifies the SPARQL triple pattern enclosed within such a request and evaluates the triple pattern using an internal storage component that contains the dataset exposed via the TPF interface. The particular storage component used by the servlet in our experiments is based on an RDF-HDT back-end~\cite{FMPGPA:13}, which manages a highly compressed, in-memory representation of RDF data.
The TPF server returns to the client an HTTP response containing the RDF triples from the evaluation of the triple pattern divided in pages~(whose size is configurable) and an estimation of the entire result set size. These result size estimations are obtained by just asking the RDF-HDT back-end how many matching triples there are for the given triple pattern. The TPF client uses such result set size estimation
	in its query execution algorithm.

\paragraph{brTPF Server Implementation:}
The brTPF server implementation extends the TPF implementation servlet by enabling it to process brTPF requests as follows. Given such a request, the servlet internally generates a stream of data triples for the requested~(brTPF) fragment and processes this stream in the same way as the TPF implementation processes the stream of matching triples returned by the RDF-HDT backend. To generate the data triples for a brTPF request, the servlet iterates over the sequence of solution mappings that comes with the request. For each such mapping, the servlet applies this mapping to the triple pattern of the request by replacing variables in the triple pattern according to the mapping. From the resulting sequence of potentially more concrete triple patterns, the servlet removes all duplicates. Next, the servlet uses each remaining triple pattern, one after the other, to query the RDF-HDT backend. The resulting streams of matching triples are then concatenated into the desired stream of all data triples for the brTPF request.



\subsection{TPF Client Implementation}
For our experiments we use a TPF client that is implemented using Node.js and that uses the TPF-based query execution algorithm for SPARQL basic graph patterns~(BGP) as proposed by Verborgh et al.~\cite{Verborgh14:LDF}. This algorithm is based on iterators that are arranged in pipelines. Query results are computed recursively by executing the pipelines. Each of these pipelines is generated for a subquery obtained from a decomposition of the initial BGP. Each iterator executes one of these subqueries returning as well an estimation of the size of its response.
The algorithm uses this estimation to adapt its execution dynamically so the subqueries with a smaller result are executed first. In this way it is possible to rapidly return a first subset of the query result.

\subsection{brTPF Client Implementation}
To develop a brTPF client we used the aforementioned Node.js-based TPF client and added a brTPF-based query execution algorithm to it. Hence, as for the servers, all basic functionality required for both client implementations is based on the same source code. This way, we ensure the comparability of our experimental results and allow for a fair comparison of
	both approaches.
For the same purpose, the brTPF-based query execution algorithm that we developed is kept deliberately simple. 
In fact, we did not attempt to integrate any sophisticated query optimization technique such as the adaptive, intermediate result based approach to generate subplans at query execution runtime as used by the TPF client. Instead, our brTPF algorithm simply choses a fixed query execution plan upfront. This plan represents a left-deep join tree that is implemented using a fixed pipeline of iterators such that each of these iterators is responsible for a different triple pattern of the query. The join order is decided based on intermediate result cardinality estimates for every triple pattern of the query. These estimates can be obtained from the server by requesting the first TPF page for each of the triple patterns.

During query execution, every iterator receives chunks of solution mappings from its predecessor. The size of these chunks corresponds to the value of \maxMpR\ as specified by the brTPF interface. Given such a chunk, the iterator issues a brTPF request consisting of the triple pattern that the iterator is responsible for and the solution mappings from the chunk. Upon arrival of the data for the requested brTPF, the iterator uses this data to generate chunks of solution mappings for the next iterator in the pipeline.

%% file: NetworkLoadExperiments.tex
\section{Experimental Comparison of Network Load}
\label{sec:NetworkLoadExperiments}

	Our first group of experiments focuses on comparing
TPF and brTPF in terms of the network load that the interfaces may cause when accessed by clients that execute SPARQL queries. In this section, we first introduce the metrics and the experimental setup
	\removable{that we} use
for these experiments, and, thereafter, we present the results of the experiments.

\subsection{Metrics}
\label{ssec:NetworkLoadExperiments:Metrics}

For this comparison we focus on two metrics: First, the \emph{number of requests}~(\numreq) that a client sends to the server during the execution of a query.
Recall that both the TPF interface and the brTPF interface split fragments into pages. Therefore, the measurements for \numreq\ do not correspond to the number of fragments~(TPF or brTPF) requested during query executions but to the number of pages requested for the fragments that the client choses to access. Notice furthermore that such requests do not necessarily have to reach the server if
	there exists
an HTTP cache
in the network between the client and the server. Nonetheless, unless the cache is located directly in front of the client~(in which case it may not be very effective), the requests are sent into the network.

	The second net\-work-re\-lat\-ed metric that we focus on is
the \emph{amount of data received}~(\dataRecv) by the client during query executions. We measure this metric in terms of the number of RDF triples contained in all fragment pages that the client receives during a query execution. Observe that this metric is independent of whether the data comes directly from the server or from an HTTP cache that acts as a proxy~server.

\subsection{Setup}
\label{ssec:NetworkLoadExperiments:Setup}
For this group of experiments we use a sin\-gle-ma\-chine setup with a single client. That is, the combined TPF/brTPF Java servlet is deployed~(using Jetty~9.2.5) on the same machine on which the client implementation performs the query executions~(using either the TPF algorithm or the brTPF algorithm). This machine runs the Ubuntu~12.04.5~LTS operating system with Oracle Java~1.8.0\_92 and Node.js~0.10.37, and it is equipped with an Intel Core i7-2620M CPU~(2.7GHz) and 8~GB of main memory.
In all of our experiments,
every query execution is performed by using a separate operating system process~(that is, we stop and restart the client software between any two executions).

As a basis for the experiments we used the 10M triples dataset%
	\ 
provided on the project page%
	\footnote{\url{http://dsg.uwaterloo.ca/watdiv/}}
of the Waterloo SPARQL Diversity Test Suite~(WatDiv)~\cite{Aluc14:WatDiv}, and we used a sequence of 145 BGP queries that we selected uniformly at random from the WatDiv stress test query workload%
	\footnote{\url{https://cs.uwaterloo.ca/~galuc/watdiv/stress-workloads.tar.gz}}%
. This workload has been shown to be a challenging benchmark that is very diverse in terms of various structural and data-driven features~\cite{Aluc14:WatDiv}.

%



\subsection{Results}
\label{ssec:NetworkLoadExperiments:Results}
For our first experiment we use a page size of 100~data triples per fragment page~(which is the default configuration of the TPF server implementation that we use as a basis for our evaluation) and execute the query sequence using the TPF client and the brTPF client, respectively. For the latter we repeat the execution of the query sequence using each of the following values for \maxMpR: 5, 10, 15, \,$\ldots$ , 45, and 50.%
\footnote{We do not increase \maxMpR\ beyond a value of 50 because the client prototype uses the HTTP method GET, which causes problems for our server if \maxMpR \textgreater 50; that is, in preliminary tests with greater values we observed server responses with status code 414~(URI~Too~Long).}
The charts in
	Figure~\ref{chart:watdiv-varyMaxMpR-numOfRequests-sum} and~\ref{chart:watdiv-varyMaxMpR-triplesReceived-sum} provide an aggregated view on
the resulting measurements for WatDiv. In particular, Figure~\ref{chart:watdiv-varyMaxMpR-numOfRequests-sum} illustrates the overall \numreq\ summed up for each client over the whole sequence of queries, respectively; similarly, Figure~\ref{chart:watdiv-varyMaxMpR-triplesReceived-sum} illustrates the sums of the \dataRecv\ measurements obtained for all queries, respectively.
%

Regarding these measurements, we make the following observations. By first focusing on brTPF only, we observe that the overall \numreq\ decreases with an increasing value for \maxMpR~(from about 131K for \maxMpR$=$5 to about 20K for \maxMpR$=$50), and so does the overall \dataRecv~(from about 1,126K for \maxMpR$=$5 to about 756K for \maxMpR$=$50).
While for \numreq\ this observation is not surprising~(if the fraction of any large intermediate result that can be sent with each request is smaller, the brTPF client has to send a greater number of such requests), 
for \dataRecv\ we explain the observation
	as follows: Each fragment page contains not only
data triples but also additional metadata triples that refer to the next and the previous page, describe the controls of the LDF interface, etc.
Therefore, when the number of fragment pages requested and received is greater~(as is the case for smaller \maxMpR; cf.\ Figure~\ref{chart:watdiv-varyMaxMpR-numOfRequests-sum}), then so is the overall number of these additional triples that have to be received with each fragment~page.

\input{ChartsNetworkLoadExperiments}

By now comparing the behavior of TPF vs.\ brTPF in the charts in Figures~\ref{chart:watdiv-varyMaxMpR-numOfRequests-sum}--\ref{chart:watdiv-varyMaxMpR-triplesReceived-sum}, we notice that for both the overall \numreq\ and the overall \dataRecv, brTPF achieves significantly smaller values.  More specifically, regarding \dataRecv, brTPF achieves between 53.5\%~(\maxMpR$=$50) and 79.6\%~(\maxMpR$=$5) of the overall \dataRecv\ of TPF~(which is about 1,414K), and for \numreq, it even goes down to
6.5\%~(\maxMpR$=$50) of the overall \numreq\ of TPF~(310K).
	At this point, we have to recall that these charts only show aggregated measurements. Hence, it might still be possible that the vastly superior behavior of brTPF as shown in these charts is actually only due to a small number of outliers.
We can verify that this is not the case by drilling into the measurements: For the different values of \maxMpR, Figure~\ref{chart:watdiv-varyMaxMpR-numOfRequests-drillIn} illustrates the number of queries for which brTPF has a smaller~(i.e., better) or greater~(i.e., worse) \numreq\ than TPF. Figure~\ref{chart:watdiv-varyMaxMpR-triplesReceived-drillIn} presents a corresponding comparison for \dataRecv.
In Figures~\ref{chart:watdiv-varyMaxMpR-numOfRequests-drillInDeeper}--\ref{chart:watdiv-varyMaxMpR-triplesReceived-drillInDeeper}, we drill in even deeper for \maxMpR$=$30~(corresponding charts for the other values of \maxMpR\ look very similar) and report the number of queries for which the difference between the \numreq~(resp.\ \dataRecv) of brTPF vs.\ TPF is between 100K to 10K, between 10K to 1K, etc. These charts show that, in terms of both \numreq\ and \dataRecv, brTPF is not only better than TPF in an impressively high number of cases, but for a large majority of these cases in which brTPF is better, the differences are significant.


To investigate whether these results are different for a different page size we conducted another experiment in which we varied the page size~(number of data triples per fragment page). That is, with both the TPF client and the brTPF client~(using \maxMpR$=$15 and \maxMpR$=$30 as exemplars), we repeated the execution of the query sequences for each of the following page sizes: 100, 250, 500, 1000, and 2000.
Due to space limitations, we do not include charts for this experiment in this paper. However, we highlight that the measurements obtained by this experiment show for both brTPF and TPF, the page size does not have any considerable impact on \numreq\ or on \dataRecv. In other words, the relative differences between brTPF and TPF as identified by the first experiment are independent of the page size~(and so are the relative differences between the different \maxMpR\ configurations for brTPF). Hence, our main conclusion from these experiments is that, \emph{independent of the page size~(and the value of \maxMpR), brTPF typically achieves a significantly smaller \numreq\ and \dataRecv~than~TPF}.

%% file: ChartsNetworkLoadExperiments.tex
\begin{figure}[t!]
\centering
	\subfigure[\footnotesize sum of all \numreq\ for the WatDiv~runs]{%
		\includegraphics[width=0.47\textwidth]{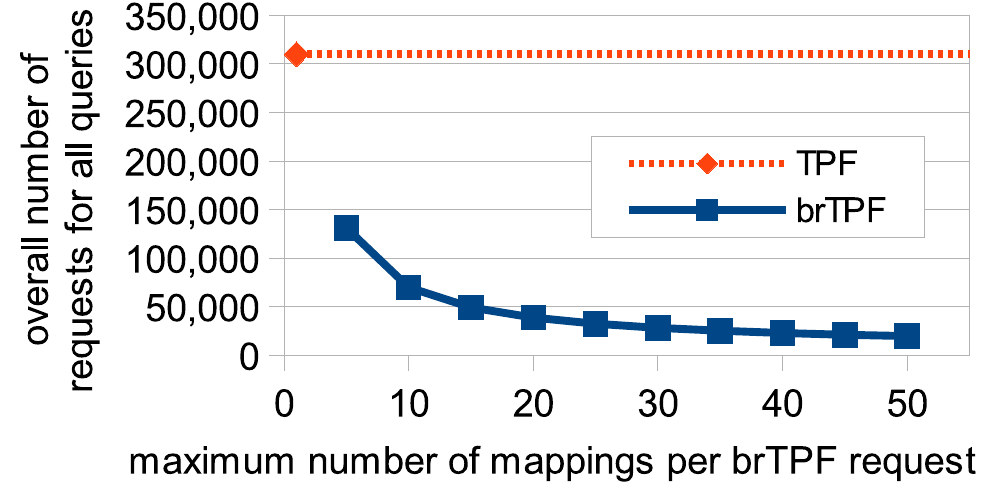}%
		\label{chart:watdiv-varyMaxMpR-numOfRequests-sum}%
	}\hfill
	\subfigure[\footnotesize sum of all \dataRecv]{%
		\includegraphics[width=0.47\textwidth]{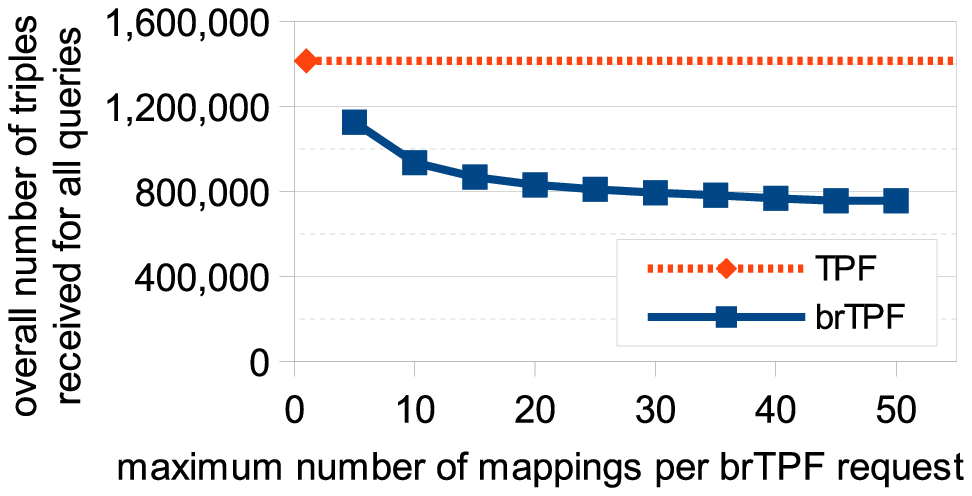}%
		\label{chart:watdiv-varyMaxMpR-triplesReceived-sum}
	}

	\subfigure[\footnotesize number of queries for which brTPF has a better/same/worse \numreq\ than~TPF]{%
		\includegraphics[width=0.47\textwidth]{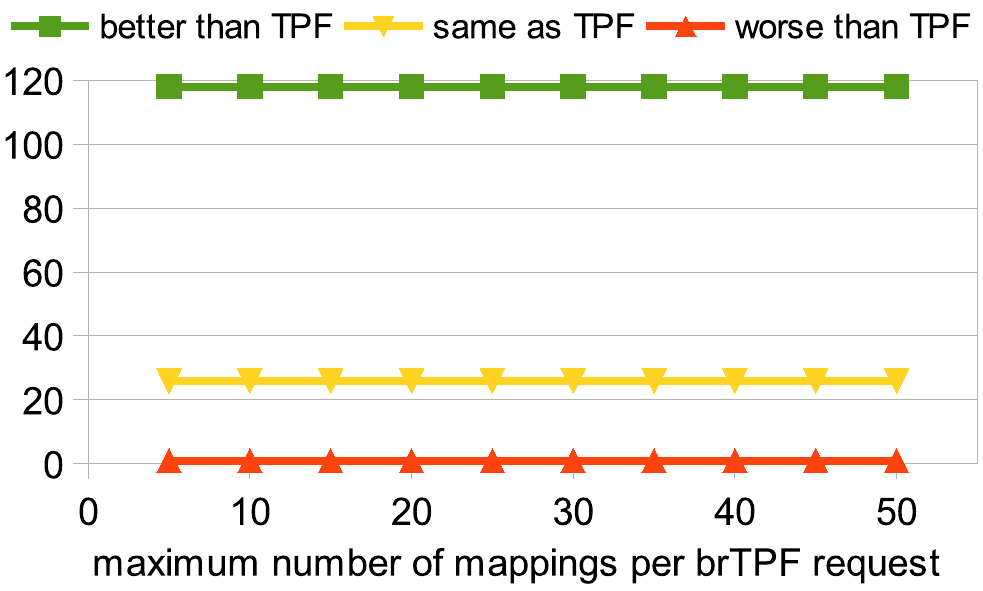}%
		\label{chart:watdiv-varyMaxMpR-numOfRequests-drillIn}%
	}\hfill
	\subfigure[\footnotesize number of queries where brTPF has a better/same/worse \dataRecv\ than~TPF]{%
		\includegraphics[width=0.47\textwidth]{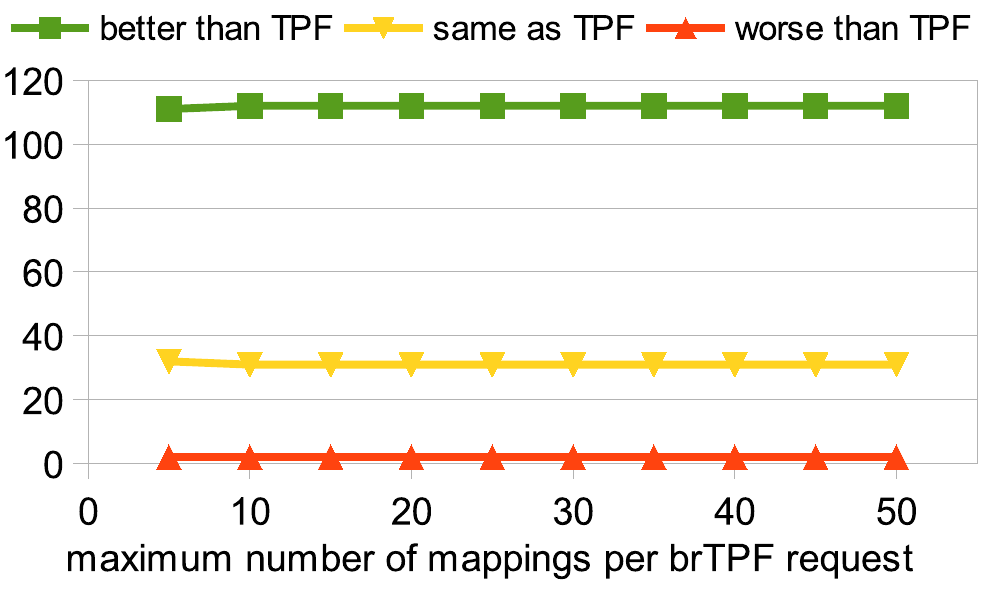}%
		\label{chart:watdiv-varyMaxMpR-triplesReceived-drillIn}
	}

	\subfigure[\footnotesize breakdown of the number of queries in terms of the differences between \numreq\ of brTPF~(\maxMpR$=$30) and of TPF]{%
		\includegraphics[width=0.47\textwidth]{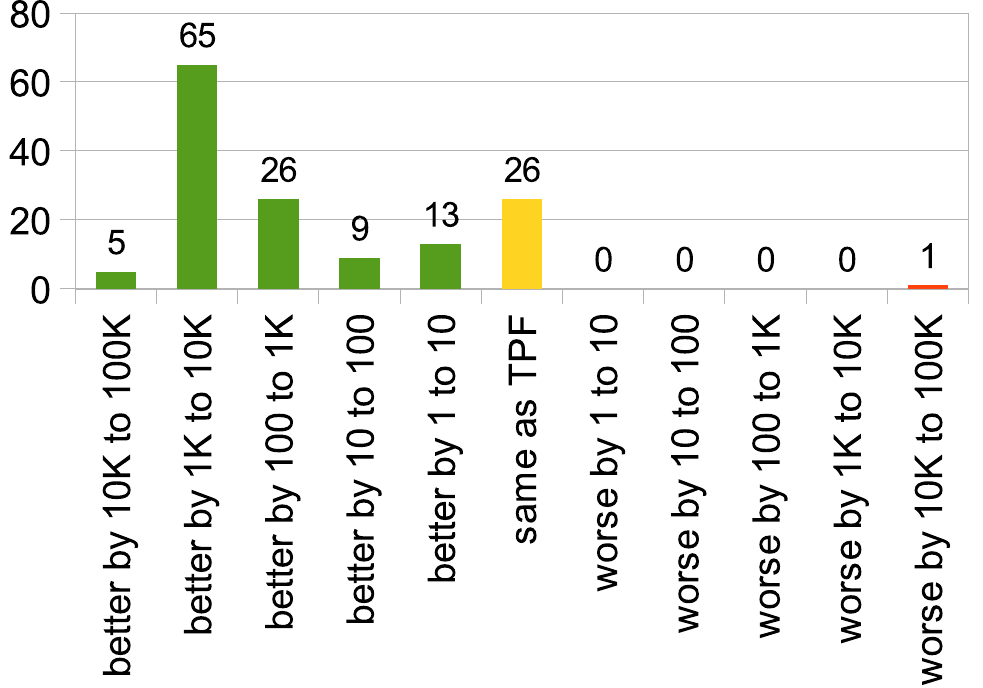}%
		\label{chart:watdiv-varyMaxMpR-numOfRequests-drillInDeeper}%
	}\hfill
	\subfigure[\footnotesize breakdown of the number of queries in terms of the differences between \dataRecv\ of brTPF~(maxM/R$=$30) and of TPF]{%
		\includegraphics[width=0.47\textwidth]{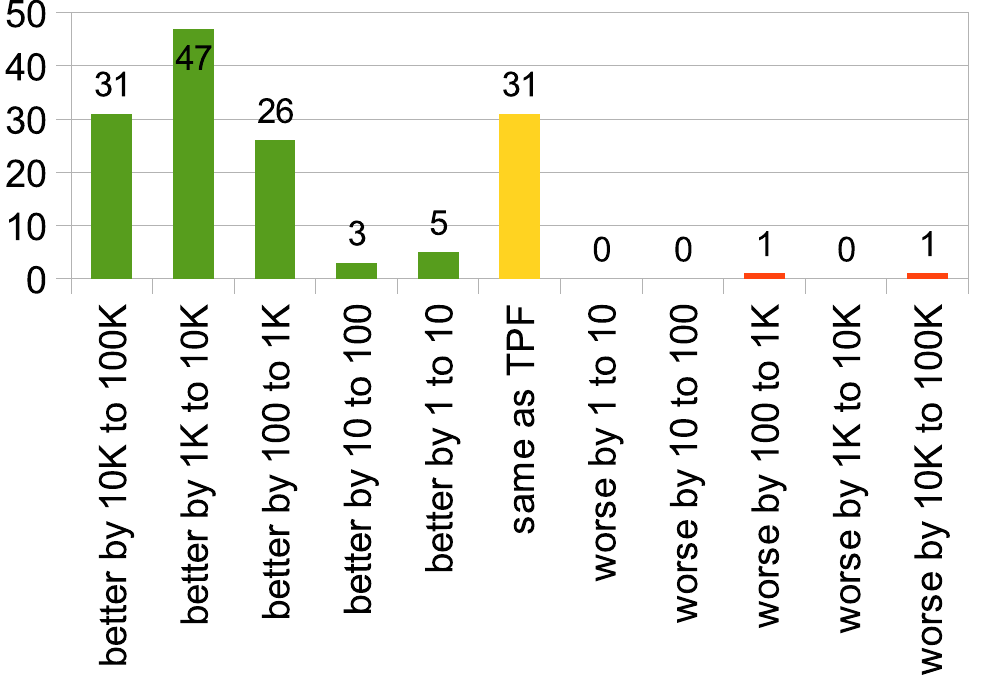}%
		\label{chart:watdiv-varyMaxMpR-triplesReceived-drillInDeeper}
	}
	\caption{Measurements of network-related metrics using WatDiv.}
	\label{charts:watdiv-varyMaxMpR-network}
\end{figure}

%% file: ServerLoadExperiments.tex
\section{Experimental Comparison of Performance under Load}
\label{sec:ServerLoadExperiments}

With our second group of experiments we aim to compare the behavior of TPF and brTPF when multiple clients access a server concurrently,%
\footnote{We also tried to conduct a comparison between TPF/brTPF and a Virtuoso-based SPARQL endpoint. Unfortunately, as detailed in%
	\ProceedingsVersionOnly{~\cite[Appendix~A]{ExtendedVersion}, }%
	\ExtendedVersionOnly{\ Appendix~\ref{appendix:Virtuoso}, }%
the Virtuoso server failed to correctly serve multiple concurrent clients in our experimental setup.}
and
	\removable{we aim to analyze}
how the two approaches scale to an increasing number of concurrent clients. To this end, we deliberately ignore the possibility of having an HTTP cache that may reduce the server load~(we consider caching in our third group of experiments as presented in Section~\ref{sec:CacheExperiments}).

\subsection{Setup}
\label{ssec:ServerLoadExperiments:Setup}
For our multi-cli\-ent experiments we use a cluster of 17~identical machines that are connected via a 10~Gb Ethernet network~(Dell PowerConnect 6248 switch). Each machine has an Intel Core~i7 processor with 4~cores at 2.6Ghz and 8~GB of main memory. The operating system running on all machines is an Ubuntu~14.04~LTS.

One of these machines we used as the server and deployed the combined TPF/brTPF Java servlet on an Apache Tomcat~8 application server running with Java~1.8. The page size in this setup we fix to 100%
	\removable{~data triples per fragment~page}%
.

The other 16~machines we used to simulate clients. For this purpose, we configured each of these
machines to run on each of its 4~cores a thread with a single brTPF/TPF client, which gives us a total of up to 64~brTPF/TPF clients that we used for executing different WatDiv query sequences in parallel. To this end, we used the aforementioned WatDiv stress test query workload which consists of a set of 12,400~different queries; we split this set into 64~distinct sets, and distributed these sets over our experiment cluster such that each of the 64 CPU cores
had available a total of 193~queries that they always executed as a sequence in the same order. We also configured each client to terminate any query execution that did not complete within 5~minutes. These query executions were recorded in our experiments as 'timed out.' Once a client stopped a query execution after 5~minutes, the client starts executing the next query from its~sequence.

\subsection{Metrics}
\label{ssec:ServerLoadExperiments:Metrics}

The
metric that we consider to study the performance of the approaches under load is \emph{query throughput}~(\throughput). We measure \throughput\ in terms of the overall number of queries that all concurrent clients manage to execute within one hour%
	\ProceedingsVersionOnly{. Additionally, we measured the query execution times and provide a detailed presentation of these measurements in Appendix~B~\cite{ExtendedVersion}. }%
	\ExtendedVersionOnly{.\footnote{Appendix~\ref{appendix:QET} complements our presentation of throughput-related results by also drilling in to the query execution times measured during our multi-cli\-ent experiments.} }%
%
%

\subsection{Results}
\label{ssec:ServerLoadExperiments:Results}

The
	three charts on the left hand side of Figure~\ref{charts:watdiv-cluster-throughput}
illustrate the measurements obtained by the experiment~(ignore the right-hand-side figures for the moment%
).

\input{ChartsServerLoadExperiments}

Before we focus on the \throughput\ values measured, we need to mention that we observed a varying, but often very high number of timed out queries across all runs. More specifically, as illustrated in Figure~\ref{chart:watdiv-cluster-noCache-timeouts}, for both TPF and brTPF, the number of timeouts increases with an increasing number of clients; and in all cases, the TPF clients run into a substantially higher number of timeouts than the brTPF clients.
We identified the following cause of this behavior: Both client implementations asynchronously issue multiple HTTP requests in parallel. The TPF server has comparably less work to do to answer each of these multiple HTTP requests, returning the data as fast as possible. However both clients at this stage still have to do some data processing to join intermediate
	results.
In the case of TPF, this client-side processing is significantly more work due to the larger amounts of data that the TPF client receives from the TPF server.
	As a result of
this extra work that the TPF client has to do, there are more query executions that exceed the 5~mins timeout threshold before they~complete.


Given the high amount of timeouts, it is important to report not only the \throughput\ of all query execution attempts, including the ones the timed out, but also the \throughput\ in terms of query executions that terminated normally after computing their complete query result. Therefore, Figure~\ref{chart:watdiv-cluster-noCache-throughputInclTimeouts} illustrates the former and Figure~\ref{chart:watdiv-cluster-noCache-throughputWithoutTimeouts} the latter.
In these charts we make two main observations: First, the brTPF approach achieves a greater \throughput\ than TPF, and, second, both approaches are able to achieve higher \throughput\ values for an increasing number of clients. However, regarding the latter we also observe that the brTPF approach scales better if we look at the \throughput\ of completed query executions, whereas, by comparing only the \throughput\ in terms of all query execution attempts, both approaches seem to exhibit a similar scaling behavior. 

As a last observation regarding this group of experiments we note that the average query execution time~(average QET) across all WatDiv queries that were executed completely~(no timeouts) in the 4-clients setup
was 17.9~secs for TPF~(st.dev.\ 33.2) and 16.5~secs for brTPF~(st.dev.\ 44.9). For the 64-clients setup this average QET increased to 45.3~secs for TPF~(st.dev.\ 61.0) and to \removable{a much smaller} 33.1~secs for brTPF~(st.dev.\ 57.7).
	\ProceedingsVersionOnly{Appendix~B in the extended version of this paper drills into these numbers and discusses QETs in more detail~\cite{ExtendedVersion}. }%
	\ExtendedVersionOnly{Appendix~\ref{appendix:QET} drills into these numbers and discusses QETs in more detail. }%

%% file: ChartsServerLoadExperiments.tex
\begin{figure}[t!]
\centering
	\subfigure[\footnotesize throughput \textbf{without cache}~(incl.\ timeouts)]{%
		\includegraphics[width=0.47\textwidth]{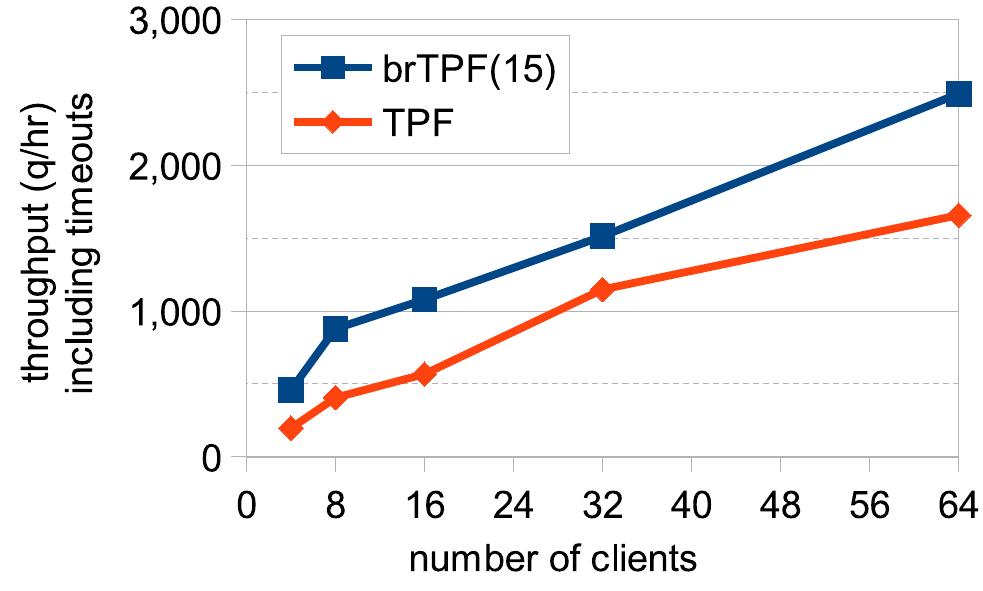}%
		\label{chart:watdiv-cluster-noCache-throughputInclTimeouts}%
	}\hfill
	\subfigure[\footnotesize throughput \textbf{with a cache}~(incl.\ timeouts)]{%
		\includegraphics[width=0.47\textwidth]{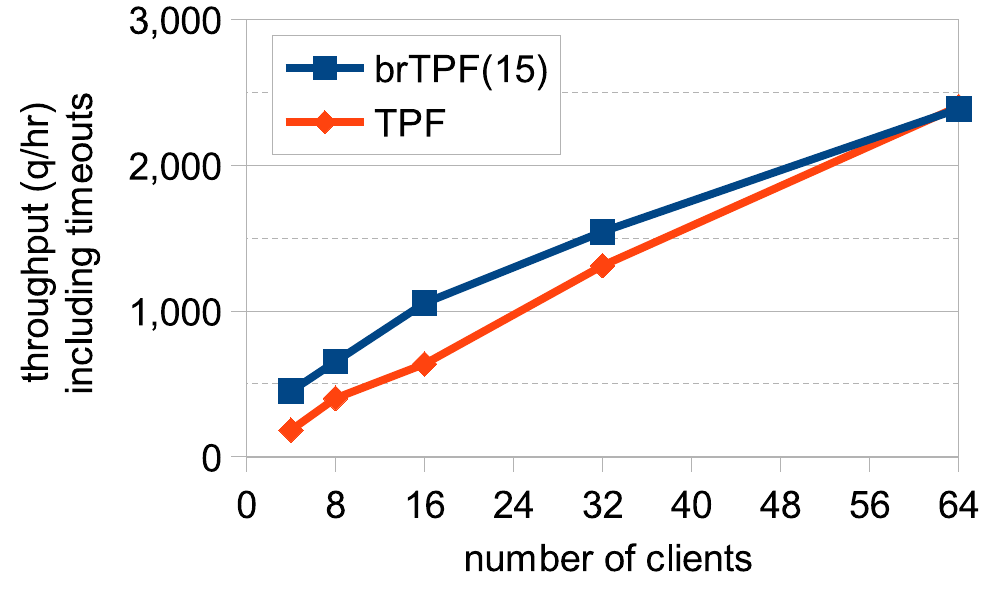}%
		\label{chart:watdiv-cluster-withCache-throughputInclTimeouts}
	}


	\subfigure[\footnotesize timed-out queries \textbf{without a cache}]{%
		\includegraphics[width=0.47\textwidth]{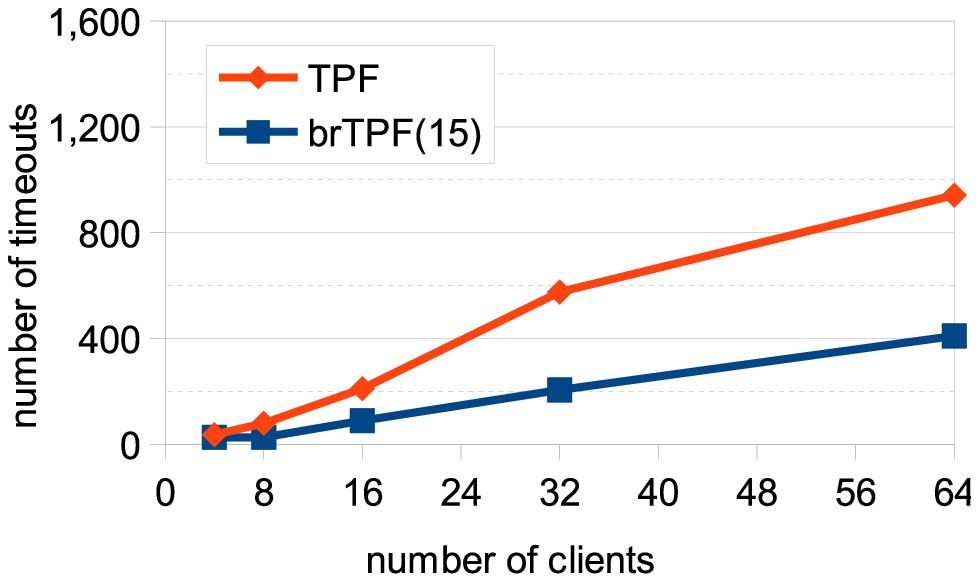}%
		\label{chart:watdiv-cluster-noCache-timeouts}%
	}\hfill
	\subfigure[\footnotesize timed-out queries \textbf{with a cache}]{%
		\includegraphics[width=0.47\textwidth]{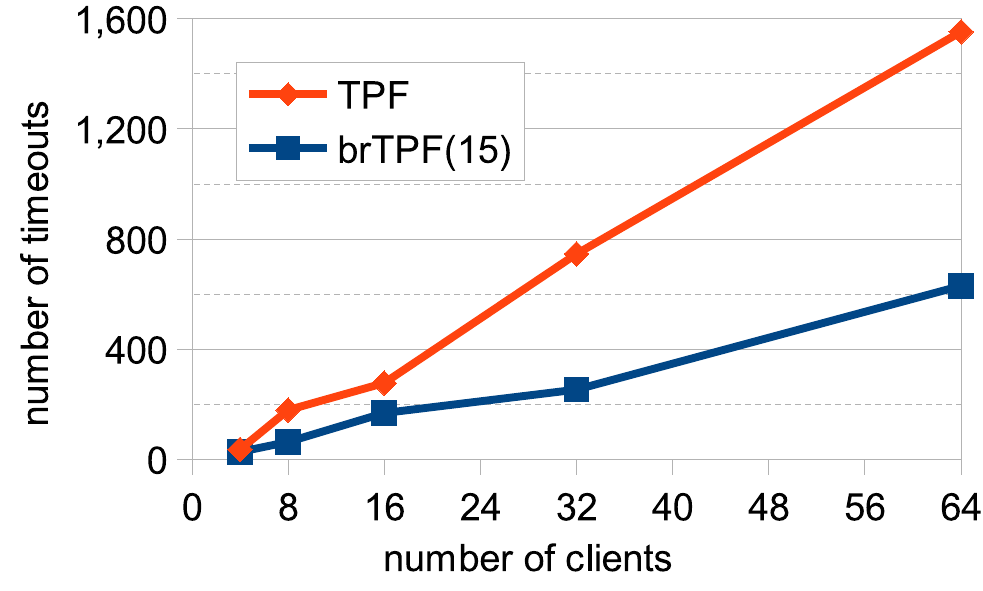}%
		\label{chart:watdiv-cluster-withCache-timeouts}
	}

	\subfigure[\footnotesize throughput \textbf{without cache}~(w/o timeouts)]{%
		\includegraphics[width=0.47\textwidth]{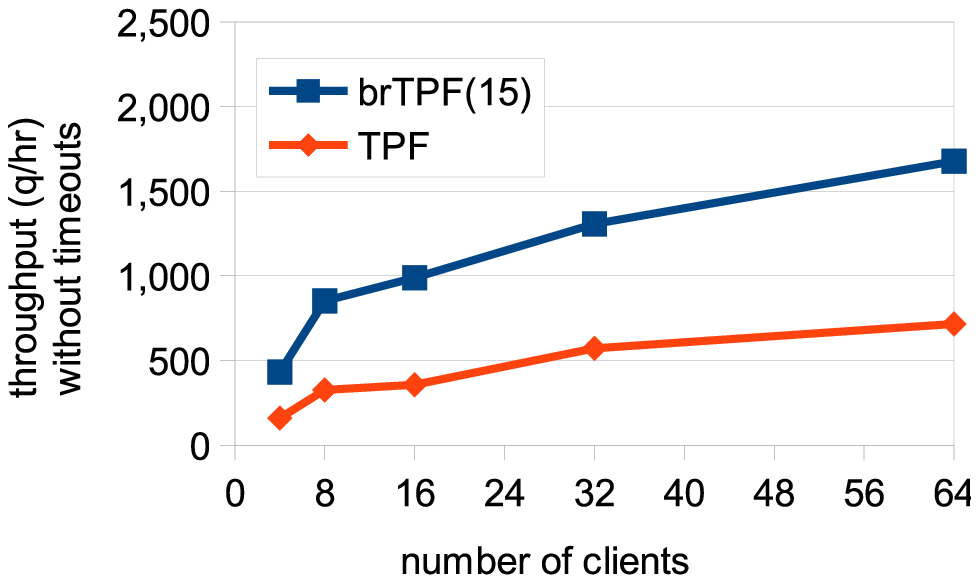}%
		\label{chart:watdiv-cluster-noCache-throughputWithoutTimeouts}%
	}\hfill
	\subfigure[\footnotesize throughput \textbf{with a cache}~(w/o timeouts)]{%
		\includegraphics[width=0.47\textwidth]{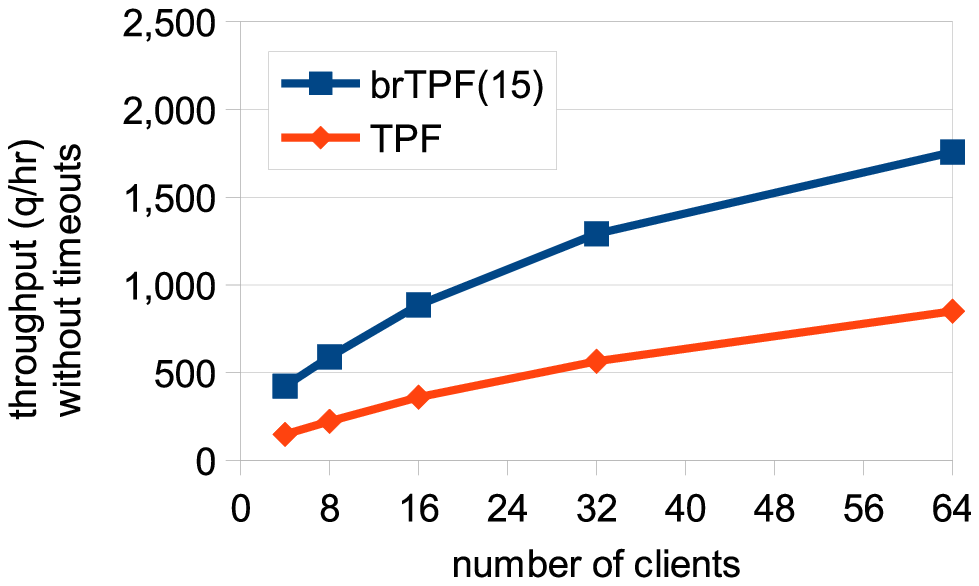}%
		\label{chart:watdiv-cluster-withCache-throughputWithoutTimeouts}
	}

	\ProceedingsVersionOnly{%
		\caption{Measurements of overall system performance in the multi-client setup.}
	}%
	\ExtendedVersionOnly{%
		\caption{Measurements of overall system performance in the multi-client setup using WatDiv.}
	}%
	\label{charts:watdiv-cluster-throughput}
\end{figure}

%% file: CacheExperiments.tex
\section{Experimental Comparison of Performance with Cache}
\label{sec:CacheExperiments}

For our analysis in the previous section we ignore the possibility of HTTP caches. We recall that such caches are designed to reduce the load of Web servers by serving requests that are identical to earlier requests~(instead of requiring the server to recompute the response for such identical requests over and over again). Therefore, our previous setup without such a cache can be conceived of as some kind of a worst-case scenario for both TPF and brTPF, and it might be worse for one than it is for the other depending on the respective likelihood for observing identical requests from different query executions. In fact, it seems reasonable to assume that this likelihood is higher for TPF than it is for brTPF. More precisely, it seems more likely that different TPF-based query executions request the same triple pattern than it is for different brTPF-based executions to request an identical pair of the same triple pattern and the same sequence of solution mappings.
%
To verify this assumption and to identify how caching affects the performance of both approaches we conducted another set of experiments. In this section, we describe these experiments and present their results.

\subsection{Cache Hit Potential}
\label{ssec:CacheExperiments:CacheHits}
To systematically study and to compare the potential for TPF requests and for brTPF requests to be served from a cache we use the same sin\-gle-ma\-chine setup as used for the net\-work-re\-lat\-ed experiments~(cf.\ Section~\ref{ssec:NetworkLoadExperiments:Setup}). As a metric for this analysis, we use the \emph{number of cache hits}~(\numhits) as could be achieved by a possible HTTP cache 
when executing the test sequence of WatDiv queries.
	We instrumented our combined TPF/brTPF server implementation to measure this number assuming either an unlimited cache or a cache whose size is limited to a given number of distinct requests~(using LRU as replacement policy).
For our analysis we first used the latter and varied the cache size from 2.5K, 5K, 10K, 50K, 100K, 250K, to~500K.

\begin{figure}[t!]
\centering
	\subfigure[\footnotesize different cache sizes (page size = 100)]{%
		\includegraphics[width=0.47\textwidth]{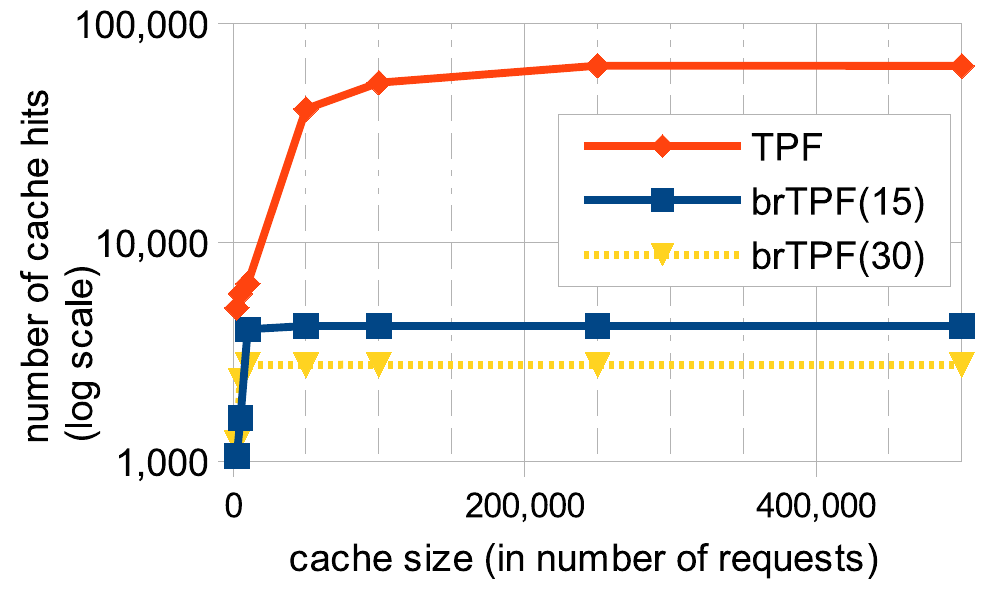}%
		\label{chart:watdiv-varyCacheSize-cachehits}%
	}\hfill
	\subfigure[\footnotesize different page sizes (unlimited cache)]{%
		\includegraphics[width=0.47\textwidth]{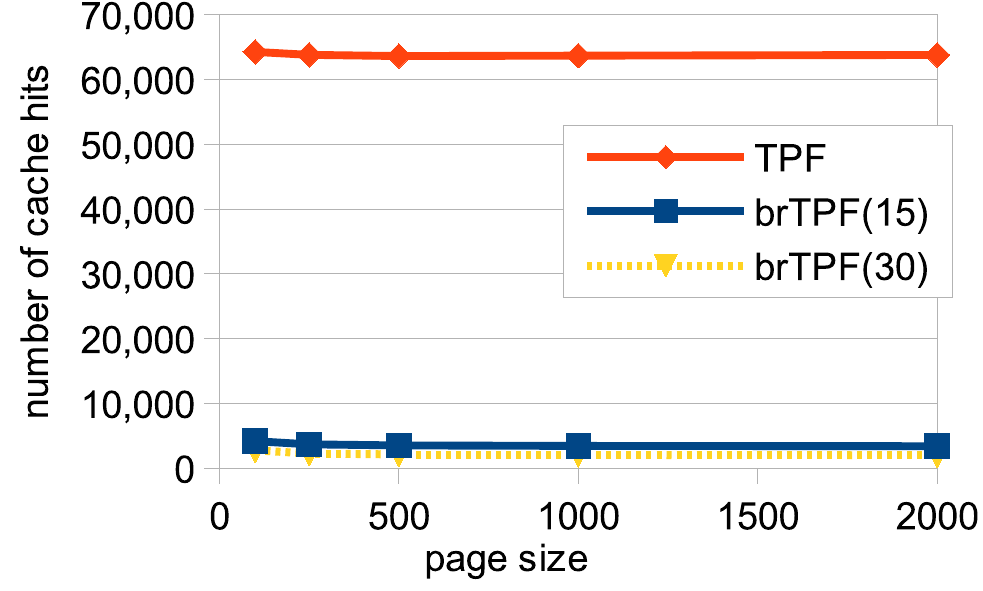}%
		\label{chart:watdiv-varyPageSize-cachehits}
	}
	\caption{Comparison of the cacheability of TPF requests vs.\ brTPF requests.}
	\label{charts:watdiv-cacheability}
\end{figure}

The chart in Figure~\ref{chart:watdiv-varyCacheSize-cachehits} illustrates our measurements for these different cache sizes when executing the WatDiv query sequence with either the TPF client or the brTPF client~(using \maxMpR$=$15 and \maxMpR$=$30 as exemplars, and a page size of 100). First and foremost, we observe that TPF always achieves a significantly higher \numhits\ than brTPF~(note that the y-axis is log scale). This observation verifies the aforementioned assumption that the likelihood for observing identical requests from different
	query executions is higher for TPF than it is for brTPF.
By focusing on brTPF, we notice that a smaller value for \maxMpR\ results in a higher \numhits. More precisely, the \numhits\ for \maxMpR$=$15 in this experiment is always about 150\% of the \numhits\ for \maxMpR$=$30, which is not surprising given that brTPF requests with
	a greater number of solution mappings \removable{attached to them} are more specialized than brTPF requests with a smaller number of solution mappings.
As a final noteworthy observation regarding the measurements in Figure~\ref{chart:watdiv-varyCacheSize-cachehits} we mention that the curves flatten out completely at some point and this point is different for each curve. The respective cache size at each of these points
	correlates with
the overall number of requests issued during the respective WatDiv run~(see the \numreq\ as reported for these runs in Figure~\ref{chart:watdiv-varyMaxMpR-numOfRequests-sum}). Unsurprisingly, for cache sizes that are large enough to cover all distinct requests
	issued during such a
run, the \numhits\ is equivalent to the \numhits\ as achieved by an unlimited cache.

In a second experiment we assumed an unlimited cache and
	executed
the query sequences with different page sizes to investigate whether the cache hit potential of both TPF and brTPF is affected by the page size. Our measurements, as reported in Figure~\ref{chart:watdiv-varyPageSize-cachehits}, show that the page size has no impact in the \numhits.
%
%
	Therefore,
%
	from these experiments we conclude
that, \emph{independent of the page size and the value of \maxMpR, TPF has a significantly higher potential to benefit from HTTP caches than~brTPF}.

\subsection{Impact of Caching on Performance}
\label{ssec:CacheExperiments:Performance}
After verifying that TPF is more likely to benefit from an HTTP cache than brTPF, the obvious question that arises is whether
	the use of a cache allows TPF to gain 
a~measurable advantage in performance. To answer this question we extended our multi-ma\-chine setup~(cf.\ Section~\ref{ssec:ServerLoadExperiments:Setup}) with an additional machine on which we run an Nginx proxy server~(1.4.6)
as an HTTP cache that is located between the client machines and the machine with the TPF/brTPF server. This additional machine has an Intel Core i7 CPU with 8~cores and 16~GB of main memory, and it
	\removable{also}
runs an Ubuntu~14.04~LTS operating system.

Given this extended setup, we repeated the same set of multi-cli\-ent executions as in the corresponding experiment without the cache~(cf.\ Section~\ref{sec:ServerLoadExperiments}). The three charts on the right hand side of Figure~\ref{charts:watdiv-cluster-throughput} illustrate the resulting measurements for the number of timeouts and the \throughput~(with and without queries whose execution timed out).

As a first observation, by comparing Figures~\ref{chart:watdiv-cluster-noCache-timeouts} and~\ref{chart:watdiv-cluster-withCache-timeouts}, we note that the number of timeouts
	for TPF has increased substantially
in comparison to the experiment without the cache.
Our explanation of such an increased number of timed out queries is similar to what we saw in Section~\ref{ssec:ServerLoadExperiments:Results}, however more exacerbated.
	That is, for many of the additional query execution attempts, the amount of data that the TPF clients receives results in client-side processing work that exceeds the 5~minutes timeout threshold.
Figures~\ref{chart:watdiv-cluster-noCache-throughputInclTimeouts} and~\ref{chart:watdiv-cluster-withCache-throughputInclTimeouts} show the query throughput by brTPF and TPF
	without a cache server and with a cache server, respectively.
By looking at Figure~\ref{chart:watdiv-cluster-withCache-throughputInclTimeouts} it is possible to see how the TPF query throughput is almost the same than brTPF's query throughput when 64~clients access the servers concurrently. This high TPF throughput is due to the large amount of timed out queries as Figure~\ref{chart:watdiv-cluster-withCache-timeouts} shows and as Figures~\ref{chart:watdiv-cluster-noCache-throughputWithoutTimeouts} and~\ref{chart:watdiv-cluster-withCache-throughputWithoutTimeouts} confirm.
By comparing
Figures~\ref{chart:watdiv-cluster-noCache-throughputWithoutTimeouts} and~\ref{chart:watdiv-cluster-withCache-throughputWithoutTimeouts}, we observe that, in the case of a high number of concurrent clients~(i.e., the 64-clients runs in our experiments), the performance \emph{of both approaches} benefits from the cache, whereas for smaller number of concurrent clients we do not observe any significant impact for either approach. To explain the latter we revisit the measurements of our sin\-gle-ma\-chine experiments: Given the \numhits\ achieved by the WatDiv runs discussed in the previous section, cf.\ Figure~\ref{chart:watdiv-varyPageSize-cachehits}, and the corresponding values for
	the overall number of requests during these runs,
cf.\ Figure~\ref{chart:watdiv-varyMaxMpR-numOfRequests-sum}, we can compute hit rates. For TPF we obtain a hit rate of about 20.7\% and for brTPF it is about 10.7\% for \maxMpR$=$15 and 11.8\% \maxMpR$=$30.%
\footnote{Although the \numhits\ is higher for \maxMpR$=$15 than for \maxMpR$=$30~(cf.\ Section~\ref{ssec:CacheExperiments:CacheHits}), the former achieves a smaller hit rate because it has a higher \numreq~(%
	\ProceedingsVersionOnly{cf.\ }%
	\ExtendedVersionOnly{as we observe in }%
Section~\ref{ssec:NetworkLoadExperiments:Results}).}
It appears that these hit rates are too small to allow for the cache to take significant load from the server in the case of a smaller number of concurrent clients. Only when the overall system becomes more busy with a higher number of concurrent clients, the availability of the cache reduces the load on the server and the throughput increases. Then, due to the greater hit rates, this increase
	\removable{in throughput} is greater for TPF than \removable{it is} for brTPF.

However, the perhaps most surprising finding is that TPF cannot benefit enough from the cache to gain an advantage over brTPF. We explain this finding by the network load that is significantly higher for TPF~(as shown in Section~\ref{sec:NetworkLoadExperiments}). In particular, the increased amounts of data that need to be transferred and processed prove to be the primary weakness of TPF, even if some of the data comes from the cache.

	In conclusion, this experiment shows that \emph{for
both approaches, \removable{TPF and brTPF}, caching helps to increase the overall performance in a multi-cli\-ent setting~(in particular for greater numbers of clients), but it does not help TPF to outperform~brTPF}.

%% file: Conclusions.tex
\section{Conclusions}
\label{sec:conclusions}

In this paper we present an interface to access RDF datasets that slightly extends the
	TPF
interface. Our extended interface, which we call Bin\-dings-Re\-strict\-ed Triple Pattern Fragments~(brTPF), allows clients to attach intermediate results to triple pattern requests. By an extensive evaluation of brTPF and TPF we obtain the following results:

\ExtendedVersionOnly{ \vspace{-2mm} } 

\begin{itemize}
\item Our main conclusion from the experimental comparison of network load is that, independent of the page size~(and the value of \maxMpR), brTPF typically achieves a significantly smaller
	number of requests 
than TPF, and less data is transferred.

\item From the experimental comparison of server performance under load our conclusions are twofold: first, the brTPF approach achieves a greater throughput than TPF, and, second, both approaches are able to achieve higher throughput values for a greater number of clients. Regarding the latter we also observe that
	brTPF
scales better if we look at the throughput in terms of completed query executions, whereas, by comparing the throughput in terms of all query execution attempts~(which includes timed-out executions), both approaches exhibit a similar scaling behavior.

\item From the experimental comparison of server performance with an HTTP cache we conclude that for both approaches, \removable{TPF and brTPF}, caching helps to increase the overall performance in a multi-cli\-ent setting~(in particular for greater numbers of clients), but it does not help TPF to outperform~brTPF.
\end{itemize}

\ExtendedVersionOnly{ \vspace{-2mm} } 
\ExtendedVersionOnly{\enlargethispage{\baselineskip}} 

In this paper we improved the idea of TPFs by significantly reducing the amount of HTTP requests and query execution times.
	Since we have shown these reductions by using a deliberately straightforward client implementation, we believe that there is even more room for improvement by using more sophisticated client-side query execution algorithms. Consequently, our future work includes investigating such algorithms.

%% file: AppendixVirtuoso.tex
	\section{Comparison to a Virtuoso-based SPARQL Endpoint}
\label{appendix:Virtuoso}

We also tried to conduct a comparison between brTPF and a Virtuoso-based SPARQL endpoint in our experimental setup.
To this end, we installed the latest version of the Open Source edition of Virtuoso~(v.7.2.4)%
	\ 
on the machine that we used as the server in our multi-machine experiments~(cf.\ Section~\ref{ssec:ServerLoadExperiments:Setup}), and we loaded the WatDiv dataset~(cf.\ Section~\ref{ssec:NetworkLoadExperiments:Setup}) into this Virtuoso endpoint.
For the clients we developed a simple Python script that submits the sequence of WatDiv queries of the respective client using the SPARQL endpoint REST API, one query after another~(recall from Section~\ref{ssec:ServerLoadExperiments:Setup} that each of the up to 64 clients uses a distinct set of 193 queries that are always executed as a sequence in the same order). Since the result cardinality of some of these queries is between 100K and 1M, to obtain complete query results and, thus, to achieve a fair comparison, we configured the result size limit of the Virtuoso server to be 1M.

For the Virtuoso-based run with 4~clients, everything went smooth and Virtuoso achieved a \throughput of 217~q/hr. With 8~clients, we saw 33~query executions hitting our 5~minutes timeout~(over the course of one hour); the throughput without these timeouts was 1,420~q/hr, which is roughly twice the throughput achieved by brTPF and roughly 4x of what the TPF approach achieved in this experiment~(cf.\ Figure~\ref{chart:watdiv-cluster-noCache-throughputWithoutTimeouts}). However, when we increased the load to 16~concurrent clients, Virtuoso always managed to answer about 300~queries before it crashed, which always happened around 2~minutes after starting this run. Before the crash, Virtuoso wrote out log messages such as the following: "System is under high load. Adding cluster nodes or using more replicated copies may needed." We provide the Virtuoso configuration file and the Virtuoso log files from this test on the Web page for this paper~(as referred to in Section~\ref{sec:intro}).

To investigate this issue we repeated the 16-clients run with a configuration in which we reduced the aforementioned result size limit to 100K. Hence, we allowed Virtuoso to return partial results. With this configuration, Virtuoso did not have any trouble serving the 16~concurrent clients, but we observed a high number of queries for which Virtuoso returned a result of the exact size of 100K. Hence, these all are incomplete query results! Therefore, from this test we conclude that the reason for why Virtuoso crashed when trying to correctly serve our 16~clients~(with complete query results) is due to the difficulty of high-cardinality WatDiv queries.

Given these
	observations,
we decided not to conduct any further comparison with the Virtuoso-based SPARQL endpoint in the context of this paper; we did not want to compare with a system that returns partial results, or crashes during the~experiments.

%% file: AppendixQueryExecTimes.tex
	\section{Query Execution Times in the Multi-Client Experiments}
\label{appendix:QET}

The average query execution time~(average QET) across all WatDiv queries that were executed completely~(no timeouts) in the 4-clients setup without cache was 17.9~s for TPF~(st.dev.\ 33.2) and 16.5~s for brTPF~(st.dev.\ 44.9). For the 64-clients setup this average QET increased to 45.3~s for TPF~(st.dev.\ 61.0) and to \removable{a much smaller} 33.1~s for brTPF~(st.dev.\ 57.7). For the setup with a cache, the relative differences are similar.

\begin{figure}[t]
\centering
	\subfigure[\footnotesize \textbf{4} clients \textbf{without a cache}]{%
		\includegraphics[width=0.47\textwidth]{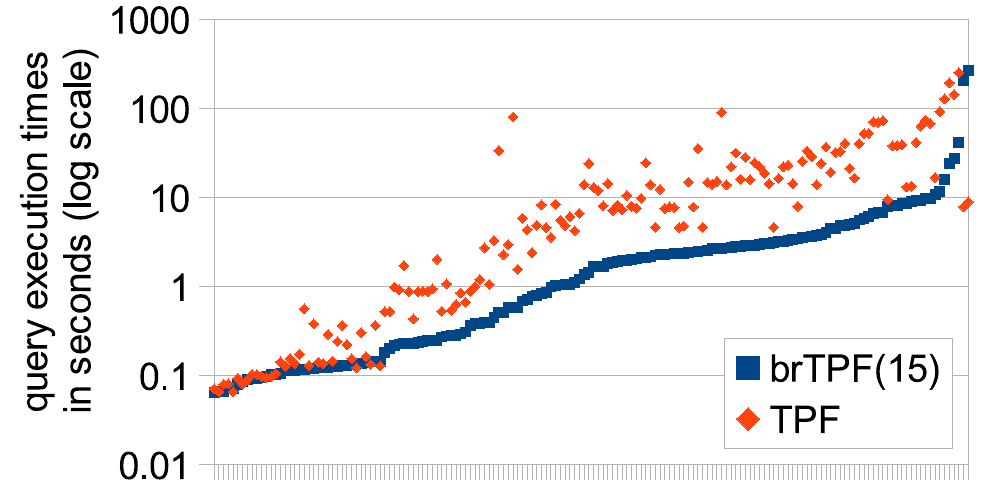}%
		\label{chart:watdiv-cluster-noCache-qexecTime-4clients}%
	}\hfill
	\subfigure[\footnotesize \textbf{4} clients \textbf{with a cache}]{%
		\includegraphics[width=0.47\textwidth]{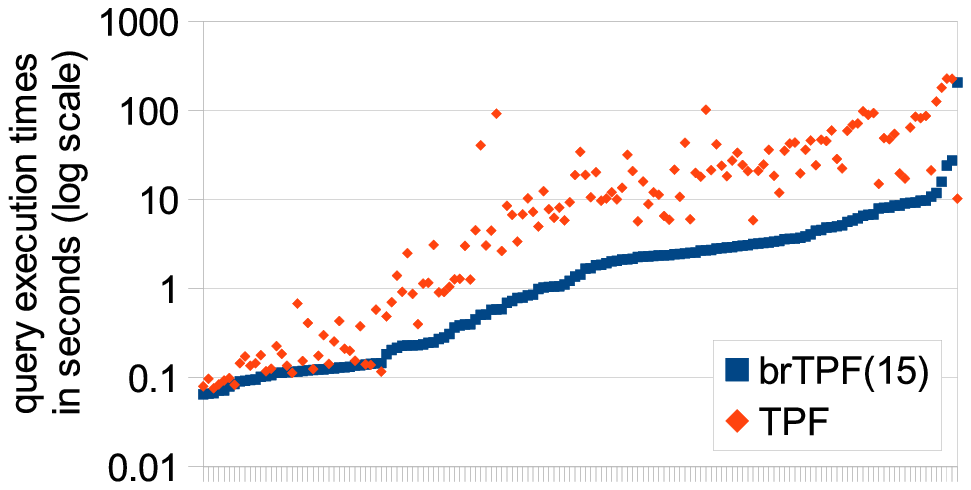}%
		\label{chart:watdiv-cluster-withCache-qexecTime-4clients}
	}

	\subfigure[\footnotesize \textbf{64} clients \textbf{without a cache}]{%
		\includegraphics[width=0.47\textwidth]{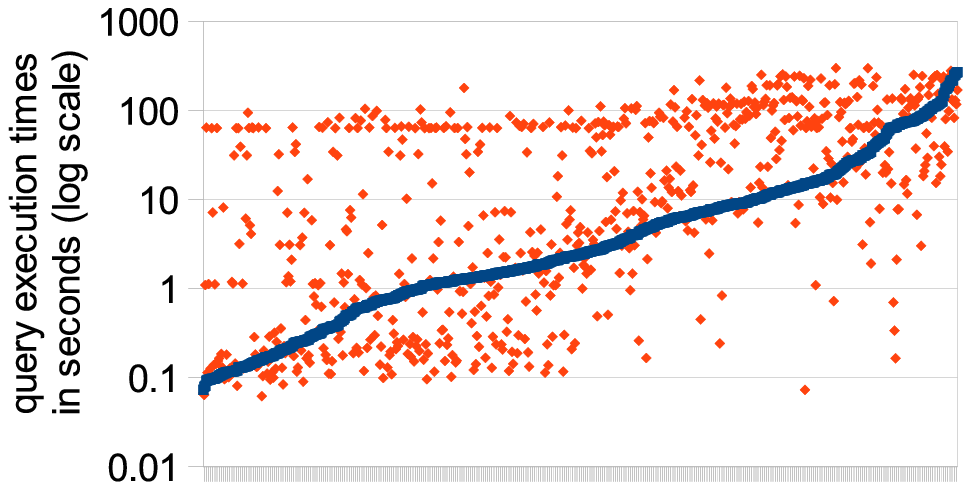}%
		\label{chart:watdiv-cluster-noCache-qexecTime-64clients}%
	}\hfill
	\subfigure[\footnotesize \textbf{64} clients \textbf{with a cache}]{%
		\includegraphics[width=0.47\textwidth]{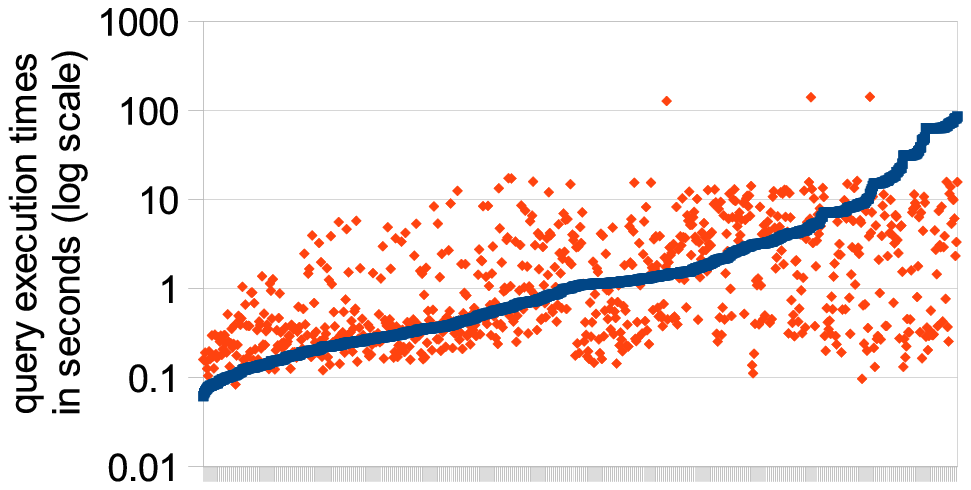}%
		\label{chart:watdiv-cluster-withCache-qexecTime-64clients}
	}
	\caption{Execution times of all the WatDiv queries that were executed completely~(without timeout) by \emph{both} the \removable{respective} TPF setup and the \removable{respective} brTPF setup~(\maxMpR=15). Hence, these charts ignore all queries that were executed completely by only one of the approaches~(due to a timeout or a smaller throughput of the other approach). In each chart the measurements are ordered from left to right by the time that brTPF required to execute the corresponding query.}
	\label{charts:watdiv-cluster-qexecTime}
\end{figure}

To drill into these numbers, the charts in Figure~\ref{charts:watdiv-cluster-qexecTime} detail the individual execution times for all queries that were executed completely by both TPF and brTPF in the respective setup.
	More precisely,
in each chart, the queries are organized along the x-axis by sorting them from left to right based on their respective QET in the brTPF case. Then, at the x-axis position of such a query, the chart contains two \removable{measurement}
	points that indicate the brTPF-based QET and the TPF-based QET of the query, respectively.
%

For the 4-clients setups without cache~(Figure~\ref{chart:watdiv-cluster-noCache-qexecTime-4clients}) and with cache~(Figure~\ref{chart:watdiv-cluster-withCache-qexecTime-4clients}) we observe that for almost all queries, brTPF achieved a QET that is either similar or
	even
smaller than the QET achieved
	by using TPF.
For the 64-clients setups~(Figures~\ref{chart:watdiv-cluster-noCache-qexecTime-64clients} and~\ref{chart:watdiv-cluster-withCache-qexecTime-64clients}), there is more variation, which we attribute to fact that, for both approaches, TPF and brTPF, individual QETs are more affected by the higher load of the whole system.
	By interpreting these charts,
the reader should keep in mind that these charts only consider queries executed completely in both the corresponding TPF and brTPF runs. What these charts do not show is that the brTPF approach achieved more than twice as many complete query executions as TPF as we recall from Figures~\ref{chart:watdiv-cluster-noCache-throughputWithoutTimeouts}~and~\ref{chart:watdiv-cluster-withCache-throughputWithoutTimeouts}.

